
\input phyzzx.tex
\date{}
\titlepage
\pretolerance=10000
\vsize 23truecm
\hsize 15.5truecm

\title{\bf Algebraic Aspects of Abelian Sandpile Models}
\author{D. Dhar$^1$, P. Ruelle$^{2,3}$, S. Sen$^4$ and
D.-\/N. Verma$^1$}

\bigskip \bigskip
\abstract{The abelian sandpile models feature a finite abelian group $G$
generated by the operators corresponding to particle addition at
various sites.  We study the canonical decomposition of $G$ as a
product of cyclic groups $G = Z_{d_1} \times Z_{d_2} \times Z_{d_3}
\cdots \times Z_{d_g}$ where $g$ is the least number of generators of
$G$, and $d_i$ is a multiple of $d_{i+1}$. The structure of $G$ is
determined in terms of the toppling matrix $\Delta$.  We construct scalar
functions, linear in height variables of the pile, that are invariant
under toppling at any site.  These invariants provide convenient
coordinates to label the recurrent configurations of the sandpile.
For an $L \times L$ square lattice, we show that $g = L$.  In this
case, we observe that the system has nontrivial symmetries,
transcending the obvious symmetries of the square, viz. those coming
from the action of the cyclotomic Galois group Gal$_L$ of the
$2(L+1)$--th roots of unity (which operates on the set of eigenvalues of
$\Delta$). These eigenvalues are algebraic integers, whose product is
the order $|G|$. With the help of Gal$_L$ we are able to group the
eigenvalues into certain subsets whose products are separately
integers, and thus obtain an explicit factorization of $|G|$. We also use
Gal$_L$ to define other simpler, though under-complete, sets of
toppling invariants.}

\bigskip
\hrule width 10cm
\smallskip
{\tentt
\noindent
$^1$ Tata Institute of Fundamental Research, Homi Bhabha Road, Bombay
400 005, India.\break
\noindent
$^2$ Departamento de F\'\i sica Te\'orica, Universidad de Zaragoza, 50009
Zaragoza, Spain.\break
\noindent
$^3$ Chercheur qualifi\'e FNRS. \hfil \break
On leave from: Institut de Physique Th\'eorique,
Universit\'e Catholique de Louvain, 1348  Louvain-la-Neuve, Belgium.\break
\noindent
$^4$ School of Mathematics, Trinity College, Dublin, Ireland.

}
\endpage

\noindent {\bf 1. {Introduction}}

The concept of self-organized criticality was proposed by Bak,
Tang and Wiesenfeld in 1987, and has given rise to growing interest in the
study of self-organizing systems.  Bak {\it et al} argued that in many natural
phenomena, the dissipative dynamics of the system is such that it drives
the system to a critical state, thereby leading to ubiquitous power law
behaviors [1,2].  This mechanism has been invoked to understand the
power--laws observed in turbulent fluids, earthquakes, distribution of
visible matter in the universe, solar flares and surface roughening of
growing interfaces [3-5].

Sandpile automata are among the simplest theoretical models which show
self--organized criticality.  A specially nice subclass consists of the
so--called abelian sandpile models (ASM's) [6].  There have been many
numerical as well as analytical studies of the ASM.  The case when there is
a preferred direction of particle transfer turns out to be equivalent to
the voter model, and all the critical exponents can be determined in all
dimensions [7].  When there is no preferred direction, the model turns
out to be related to the $q \rightarrow 0$ limiting case of the Potts
model [8].  The problem has been solved exactly in the mean field limit
[9-11].  In two and three dimensions, only some of the critical exponents
of the problem are known analytically [6,8,12].

Most of these papers have been concerned with the critical properties of
the ASM in the thermodynamic limit of large system sizes.  In this paper,
our interest is rather in the properties of the {\it finite} automata.
There have been only a handful of papers addressing these so far (and mainly
in two dimensions). Creutz' paper
[13] exhibits very interesting examples of geometrical patterns displayed
by a special configuration (the so--called identity configuration) of the ASM
on a square lattice.  Liu {\it et al} [14] have studied the patterns obtained
by relaxing some simple unstable configurations.  Wiesenfeld {\it et al} [15]
have studied the periods of deterministic ASM, again on a square lattice.
These studies have been extended by Markosova and Markos [16] to lattices
of size up to 19. This paper was inspired partly by the `experimental' results
of these authors. We will concentrate on properties such as the
structures of the abelian group and of the space of recurrent configurations.

The plan of this paper is as follows. In section 2, we give the definition of
general ASM, and review their basic properties. In section 3, we construct,
for a general ASM, a set of functions, defined on the space of recurrent
configurations, which are invariant under topplings, and which can be used to
label those configurations. With these functions, we show how to determine the
structure of the abelian group of the ASM in terms of the normal form
decomposition of its toppling matrix. In section 4, we consider the problem
of computing the rank of the abelian group when the ASM is defined on an
$L_1 \times L_2$ rectangular portion of the square lattice, and show that the
rank equals $L$ on an $L \times L$ square. In section 5, we
recapitulate briefly the results from general Galois theory needed by
us, and use them to study the Galois
group of the characteristic polynomial of the toppling matrix, and
construct, in section 6, another set of algebraic functions, invariant under
topplings, thereby extracting information about the abelian group from
Galois theory. This information is incomplete and, we show that the
method does not give the full structure of the group in general. In
section 7, we remark on two other properties of the ASM on a square
lattice. Firstly, we derive a sharper upper bound on the
time period of the deterministic ASM studied in [15]. The second one is an
interesting connection between the structure of identity configurations on $2L
\times 2L$ and $(2L + 1) \times (2L + 1)$ lattices. Some technical material
involving detailed calculations is given in appendices A and B, while
in appendix C we display some identity configurations for some small
square lattices. In appendix D, we show how the toppling
invariants discussed by Lee {\it et al} [17] are a special case of the
invariants discussed in this paper.

\bigskip
\bigskip

\noindent {\bf 2. Preliminaries and notation}

The general ASM is defined on a set of $N$ sites, labelled say by integers 1
to $N$.  Each site $i$ is assigned an integer variable $z_i$, called the
height of the sandpile at site $i$.  The time evolution of the sandpile is
defined in terms of the following two processes:
\medskip

\item{(1)} Addition of particles: We choose a site at random, and increase
its height by 1, while the heights at other sites remain unchanged.
The probability of choosing the $k^{\rm th}$ site is denoted by $p_k$. For
simplicity we take all $p_k$'s nonzero. [This
condition is needed to ensure the uniqueness of the steady state.]

\item{(2)} Relaxation: If the height at some site $j$ equals or exceeds a
prespecified value $z_{crit} (j)$, it is said to be unstable and
topples, loosing some grains of sand which either fall on other
`neighbouring' sites (whose height increases as a consequence), or
drop out of the system.  The updating of heights is specified in terms
of an $N \times N$ integer matrix $\Delta$, called the toppling matrix and
satisfying $\Delta_{ii} \geq 0$ and $\Delta_{ij} \leq 0$ for $i \neq j$.
If there is a toppling at some site $j$, all heights $\{z_i\}$ are updated
according to the rule
$$ {\rm If}\;\;z_j \geq z_{crit} (j), \;\;\;{\rm then}\;\; z_i
\rightarrow z_i - \Delta_{ij} \hbox{  for all } i.
\eqno (2.1)
$$
We may assume that
$$
z_{crit} (j) = \Delta_{jj},
\eqno (2.2)
$$
so that the values of $z_j$ in a stable configuration are between $0$ and
$\Delta_{jj}-1$.

A toppling at one site may make other sites unstable.  A sequence of
topplings
caused by adding a single particle is called an avalanche.  When all the
unstable sites have toppled, we are left with a new stable configuration.
This defines a single step of time evolution. At the next time, a new
particle is added at a random site, and the system is allowed to
relax, and so on.

It is convenient to define operators $a_k$, $k=1$ to $N$,
corresponding to the process of adding a particle at site $k$, and allowing
the system to relax.  These operators $a_k$ map stable configurations into
stable configurations.

A general analysis of abelian sandpile models was carried out in [6].  It
was shown that the operators $\{a_k\}$ commute with each others.  This
allows a simple characterization of the critical steady state: Only a
small subset of all stable configurations occur with non--zero probability
in the steady state. These are called recurrent configurations. Their
number is equal to Det$\,\Delta$, and in the steady state they occur
with equal probability.
The $a_k$'s map the space $\cal R$ of recurrent configurations onto
itself, and are invertible on $\cal R$.

Let $G$ be the abelian group generated by the operators $\{a_i\}$, $i=1$ to
$N$. This is a finite group as these operators satisfy the closure relations
[6] $$
\prod^N_{i=1} a_i^{\Delta_{ij}} = I, \hbox{ for all }j.
\eqno (2.3)
$$
The order of $G$, denoted by $|G|$, is equal to the number of
recurrent configurations. This is a consequence of the fact that if
$C$ and $C'$ are any two recurrent configurations, then there is an
element $g \in G$ such that $C'=gC$. We thus have
$$ |G| = |{\cal R}| = {\rm Det}\,\Delta.
\eqno (2.4)
$$

\bigskip
\bigskip

\noindent {\bf 3. Toppling invariants and the group structure for general
ASM}

The space of all configurations $\{z_i\}$ (with non--negative heights
$z_i$) constitutes a commutative semigroup over the given set of
$N$ sites, with the operation given by sitewise addition of heights followed
by relaxation if necessary. One can define an equivalence relation on this
semigroup by declaring two configurations $\{z_i\}$ and $\{z_i^\prime\}$
equivalent (under toppling) if and only if there exist $N$ integers
$n_j$ such that
$$ z_i^\prime \;=\; z_i \;-\; \sum_j \; \Delta_{ij} \, n_j\;,
\qquad \hbox{for all $i$}.
\eqno (3.1)
$$
Each equivalence class contains one and only one recurrent configuration.
Indeed to any configuration $\{z_i\}$, one can associate a recurrent
configuration $C$ by letting
$$ C[z_i] = \prod_i \, a_i^{z_i} \,C^*,
\eqno(3.2)
$$
where $C^*$ is any fixed recurrent configuration. Then if $\{z_i\}$ and
$\{z_i^\prime\}$ are related as in (3.1), one easily checks that $C[z_i]=
C[z'_i]$ on account of the relations (2.3). (Note that with the
definition (3.1), two stable configurations may be equivalent under toppling.)

Toppling invariants are scalar functions defined on the space of all
configurations of the sandpile, such that they take the same value for any
two configurations which are equivalent under toppling.
Toppling invariants which are linear in the height variables, and which are
conserved modulo various integers were first introduced by Lee and Tzeng [17]
in the context of specific deterministic one--dimensional ASM. Their results
will be shown to be particular cases of our general construction (see
Appendix D).

Given the toppling matrix $\Delta$, we define $N$ rational functions
$Q_i$ ($i = 1$ to $N$) by setting
$$
Q_i (\{z_j\}) = \sum_j \Delta^{-1}_{ij}\,z_j \;\;\bmod 1.
\eqno (3.3)
$$
It is straightforward to prove that the functions $Q_i$ are toppling
invariants. Indeed, under toppling at
site $k$, the configuration $C \equiv \{z_j\}$
changes to $C^\prime \equiv \{z_j^\prime = z_j - \Delta_{jk}\}$, and
from the linearity of the functions $Q_i$ in the height variables, we have
$$
Q_i(C') = Q_i(C) - \sum_j \Delta^{-1}_{ij} \Delta_{jk} = Q_i(C) \;\;\bmod 1.
\eqno (3.4)
$$
The functions $Q_i$ take rational values, but they are easily made
integer--valued upon multiplication by some adequate integer.
Being toppling invariants, the functions $Q_i$ can be used to label
the recurrent configurations, and thus the space $\cal R$ can be
replaced by the set of $N$--uples $(Q_1,Q_2,\ldots,Q_N)$. However, the
labelling by the $Q_i$'s is generally overcomplete, they not being all
independent. A simple example will readily establish that.

Throughout this
paper, we shall consider in detail the special case of the ASM defined on an
$L_1 \times L_2$ rectangular portion of a two--dimensional square lattice.
We choose the toppling matrix to be the discrete Laplacian, whose diagonal
entries are given by $\Delta_{ii} = 4$, and the off-diagonal entries
$\Delta_{ij} = -1$ or $0$ according to whether the sites $i$ and $j$ are
nearest neighbours or not. Note
that this implies open boundary conditions on all four boundaries of the
rectangle, and thus any toppling there involves a loss of sand.

To be specific, consider the case $L_1 = L_2 = 2$, with the configurations
specified as $\pmatrix{z_1 & z_2 \cr z_3 & z_4}$. In this
case $\Delta$ is a $4 \times 4$ matrix of determinant $192$, and we find
$$\Delta = \pmatrix{4 & -1 & -1 & 0 \cr -1 & 4 & 0 & -1 \cr
                   -1 & 0 & 4 & -1 \cr 0 & -1 & -1 & 4 \cr}
\quad {\rm and} \quad
\Delta^{-1} = {1 \over 24} \pmatrix{7 & 2 & 2 & 1 \cr 2 & 7 & 1 & 2 \cr
                                   2 & 1 & 7 & 2 \cr 1 & 2 & 2 & 7 \cr}.
\eqno(3.5)
$$
The definition (3.3) yields the four invariants
$$\eqalignno{
Q_1 &= {1 \over 24}(7z_1+2z_2+2z_3+z_4) \;\;\bmod 1, &(3.6a) \cr
Q_2 &= {1 \over 24}(2z_1+7z_2+z_3+2z_4) \;\;\bmod 1, &(3.6b) \cr
Q_3 &= {1 \over 24}(2z_1+z_2+7z_3+2z_4) \;\;\bmod 1, &(3.6c) \cr
Q_4 &= {1 \over 24}(z_1+2z_2+2z_3+7z_4) \;\;\bmod 1. &(3.6d) \cr}
$$
They satisfy three linear relations
$$\eqalignno{
& Q_1 + Q_2 = 0 \;\;\bmod 1/8, &(3.7a) \cr
& Q_3 = 4Q_1 - Q_2 \;\;\bmod 1, &(3.7b) \cr
& Q_4 = - Q_1 + 4Q_2 \;\;\bmod 1, &(3.7c) \cr}
$$
from which one sees that only two independent invariants remain, which we
choose integer--valued for convenience
$$\eqalignno{
I_1 &\equiv 24Q_1 = 7z_1+2z_2+2z_3+z_4 \;\;\bmod 24, &(3.8a) \cr
I_2 &\equiv 8(Q_1 + Q_2) = 3z_1+3z_2+z_3+z_4 \;\;\bmod 8.
&(3.8b) \cr}
$$
These two invariants provide a complete labelling for the space of recurrent
configurations, of cardinality $192$.

As this very simple example already shows, for an arbitrary $N \times N$
matrix $\Delta$, we construct $N$ toppling invariants $Q_i$, but they are
generally not independent. It thus seems desirable to isolate a minimal
set of independent invariants, as we did above to obtain the mininal set
given by $(I_1,I_2)$. We now show how this can be done for an
arbitrary ASM using the classical theory of Smith normal form for
integer matrices.

In current mathematics literature, as in Jacobson [18], this is often
discussed for matrices with entries in a principal ideal domain (the ring of
integers being a prime example). From Theorems 3.8 and 3.9 of Jacobson,
any nonsingular $N \times N$ matrix $\Delta$ can be written
in the form
$$
\Delta = ADB,
\eqno (3.9)
$$
where $A$ and $B$ are $N \times N$ integer matrices with determinants $\pm
1$, and $D$ is a diagonal matrix
$$
D_{ij} = d_i \delta_{ij}
\eqno (3.10)
$$
such that
\item{1.} $d_i$ is a multiple of $d_{i+1}$ for all $i = 1$ to $N-1$, and
\item{2.} $d_i = e_{i-1}/e_i\;$, where $e_i$ stands for the greatest common
divisor of the determinants of all the $(N-i) \times (N-i)$ submatrices
of $\Delta$ (set $e_N=1$).

\noindent
Thus the matrix $D$ is uniquely determined by $\Delta$, but the
matrices $A$ and $B$ are far from unique. The integers $d_i$ are
are called the elementary divisors of $\Delta$. An efficient algorithm to
compute them can be found in Cohen [19].

In terms of the decomposition (3.9), we define the set of scalar
functions $I_i(C)$ by
$$
I_i(C) = \sum_j (A^{-1})_{ij}z_j \quad \bmod d_i.
\eqno (3.11)
$$
Due to the unimodularity of $A$, these functions are integer--valued.
As argued for the $Q$'s in Eq. (3.4), we see using (3.9) that $I_i(C)$
so defined are invariant under the toppling of any site. Clearly, only
those invariants $I_i$ with $d_i \neq 1$ are nontrivial. Note that
they are written in terms of $A$, and hence not unique. It will be
obvious from the discussion below that the set of non--trivial $I_i$
is always minimal and complete.

As suggested by the notation used in the above example, this
second set $\{I_i\}$ is a minimal set of invariants chosen from the
overcomplete set $\{Q_i\}$. The Smith decomposition precisely shows
how to combine the overcomplete invariants $Q_i$ so as to obtain a
complete set. Indeed, it is easily checked that the $I_i$'s can be
written in terms of the $Q_i$'s as
$$I_i = \sum_j d_i \,B_{ij}Q_j \;\;\bmod d_i.
\eqno(3.12)
$$
For instance, in the above $2 \times 2$ example, one finds the following Smith
decomposition (written in the form $A^{-1}\Delta = DB$)
$$\pmatrix{7&2&2&1 \cr 3&3&1&1 \cr -1&0&0&0 \cr -2&0&-1&0 \cr}\;\Delta \;=\;
\hbox{diag$\,(24,8,1,1)$}\; \pmatrix{1&0&0&0 \cr 1&1&0&0 \cr -4&1&1&0 \cr
-7&2&-2&1 \cr}.
\eqno(3.13)
$$
 From the first two rows of $A^{-1}$ and (3.11), one recovers the two
invariants $I_1,\,I_2$ of (3.8). From the matrix $B$ and (3.12), one finds
the relations (3.8) between $(I_1,I_2)$ and $(Q_1,Q_2)$, while the trivial
character of $I_3,\,I_4$ leads to the linear relations (3.7b--c) among the
$Q$'s.

Let us now show that the set $\{I_i\}$ not only forms a complete set of
toppling invariants, but that they also determine the structure of the abelian
group $G$.

Let $g$ be the number of $d_i > 1$. With each recurrent configuration
of the ASM, we associate a $g$--uple $(I_1, I_2, .... I_g)$, where $0 \leq I_i
< d_i$. The total number of distinct
$g$--uples is $\prod^g_{i=1} d_i = |G|$ (since $A$ and $B$
in the Smith decomposition of $\Delta$ are unimodular).

We first show that this mapping from the set of recurrent configurations to
$g$--uples is one--to--one. Let us define operators $e_i$ by the equation
$$
e_i = \prod^N_{j=1} a_j^{A_{ji}}, \qquad 1\leq i\leq g.
\eqno (3.14)
$$
Acting on a fixed recurrent configuration $C^* = \{z_j\}$, $e_i$ yields a new
recurrent configuration, equivalent under toppling to the configuration
$\{z_j + A_{ji}\}$. If the $g$--uple corresponding to $C^*$ is $(I^*_1, I^*_2,
\ldots ,I^*_g)$, it is easy to see from (3.11) that $e_i C^*$
gives a configuration whose toppling invariants are $I_k = I^*_k +
\delta_{ik}$. By operating with these operators $\{e_i\}$
sufficiently many times on $C^*$, all $|G|$ values for the $g$--uple
$(I_1, I_2, \ldots , I_g)$ are obtainable.
Thus, there is at least one recurrent configuration
corresponding to any $g$--uple $(I_1,I_2,\ldots ,I_g)$. As the total number of
recurrent configurations exactly equals the total number of $g$--uples, we see
that there is a one--to--one correspondence between the $g$--uples $(I_1, I_2,
\ldots , I_g)$ and the recurrent configurations of the ASM.

To express the operators $a_j$ in terms of $e_i$, we need to invert the
transformation (3.14). This is easily seen to be
$$
a_j = \prod_{i=1}^g e_i^{(A^{-1})_{ij}}\,, \qquad 1 \leq j \leq N.
\eqno (3.15)
$$
Thus the operators $e_i$ generate the whole of $G$. Since $e_i$ acting on
a configuration increases $I_i$ by 1, leaving the other invariants unchanged,
and since $I_i$ is only defined modulo $d_i$, we see that
$$
e_i{}^{d_i} = I, \qquad \hbox{for $i=1$ to $g$}.
\eqno (3.16)
$$
Note that the definition (3.14) for $e_i$ makes sense for $i$ between
$g+1$ and $N$, and implies relations among the $a_j$ operators ($e_i=I$ from
(3.16)).

This shows that $G$ has the canonical decomposition as a product of
cyclic groups $$
G = Z_{d_1} \times Z_{d_2} \times \ldots \times Z_{d_g},
\eqno (3.17)
$$
with the $d_i$'s defined in (3.10). We thus have shown that the generators and
the group structure of $G$ for an arbitrary ASM can be entirely determined from
its toppling matrix $\Delta$, through its normal from decomposition (3.9).

The invariants $\{I_i\}$ also provide a simple additive representation of the
group $G$. As mentioned at the beginning of this section, one can define a
binary operation of ``addition'' (denoted by $\oplus$) on the space $\cal R$
of recurrent configurations by adding heights sitewise, and then allowing the
resultant configuration to relax.
 From the linearity of the $I_i$'s in the height variables, and their
invariance
under toppling, it is clear that under this addition of configurations, the
$I_i$ also simply add, {\it i.e.} for any recurrent
configurations $C_1$ and $C_2$, one has
$$ I_i(C_1 \oplus C_2) = I_i(C_1) + I_i(C_2) \;\; \bmod d_i.
\eqno (3.18)
$$
The $I_i$'s provide a complete labelling of $\cal R$.
There is a unique recurrent configuration, denoted by
$C_{\rm id}$, for which all $I_i(C_{\rm id})$ are zero. Also, each
recurrent $C$ has a unique inverse $-C$, also recurrent, and determined by
$I_i(-C) = -I_i(C) \bmod d_i$. Therefore the addition $\oplus$ is a group law
on $\cal R$, with identity given by $C_{\rm id}$. A simple recursive algorithm
to compute $C_{\rm id}$ has been given by Creutz [13].

There exists a one--to--one correspondence between
the recurrent configurations of ASM, and the elements of the group $G$: we
associate with the group element $g \in G$, the recurrent configuration
$gC_{{\rm id}}$. It is then easy to see from (3.18) that for all $g,g' \in G$
$$
gC_{{\rm id}} \oplus g'C_{{\rm id}} = (gg') C_{{\rm id}}\,.
\eqno (3.19)
$$
Thus the recurrent configurations with the operation $\oplus$ form a
group which is isomorphic to the multiplicative group $G$, a result first
proved in [13].

The invariants $\{I_i\}$
provide a simple labelling of all the recurrent configurations. Since a
recurrent configuration can also be uniquely specified by the height variables
$\{z_i\}$, the existence of forbidden subconfigurations in ASM's implies that
these heights satisfy many inequality constraints. [Certain patterns
cannot appear inside any recurrent configuration, and hence
are called forbidden subconfigurations [6]; for instance, two neighbouring
sites can never both have zero heights.]

\bigskip
\bigskip

\noindent {\bf 4. The rank of G for a rectangular lattice}

For a general matrix $\Delta$, it is difficult to say much more about the
group structure of $G$.  In the rest of this paper, unless otherwise
specified, we shall consider the special case when $\Delta$ is the
toppling matrix corresponding to a finite $L_1 \times L_2$ rectangle of
the (two--dimensional) square lattice.
Here it is more convenient to label the sites not by a single index
$i$ going from 1 to $N=L_1L_2$, but by two Cartesian coordinates $(x,y)$,
where $1 \leq x \leq L_1$ and $1 \leq y \leq L_2$. The toppling matrix is the
discrete Laplacian, given by $\Delta(x,y;x,y)=4$,  $\Delta(x,y;x',y')= -1$
if the sites $(x,y)$ and $(x',y')$ are nearest neighbours
({\it i.e.} $|x-x'|+|y-y'|=1$), and zero otherwise.
Without loss of generality, we can assume that $L_1 \geq L_2$.

The relations satisfied by the particle addition operators $a(x,y)$ can
be written in the form
$$
a(x+1,y) = a^4(x,y) \; a^{-1}(x,y+1) \; a^{-1}(x,y-1) \; a^{-1}(x-1,y)\,,
\eqno (4.1)
$$
where we adopt the convention that
$$
a(x,0) = a(x,L_2+1) = a(0,y) = a(L_1+1,y) = I, \quad \hbox{for all $x,y$}.
\eqno (4.2)
$$
The Eqns (4.1) can be recursively solved to express any operator
$a(x,y)$ as a product of powers of $a(1,y)$. Therefore the group $G$ can be
generated by the $L_2$ operators $a(1,y)$. Denoting the rank of $G$
(minimal number of generators) by $g$, this implies that
$$
g \leq L_2\,.
\eqno (4.3)
$$
In the special case of the linear chain, $L_2=1$, we see that $g=1$, and thus
$G$ is cyclic.

 From (4.1), $a(L_1 + 1, y)$ can also be expressed as a product of powers of
$a(1,y)$, say
$$
a(L_1 + 1, y) = \prod_{y^\prime} a(1, y^\prime)^{n_{yy'}}\;,
\eqno (4.4)
$$
where the $n_{yy'}$ are integers which depend on $L_1$ and $L_2$,
and which can be explicitly determined by solving the linear recurrence
relation (4.1). The condition $a(L_1+1,y)=I$ then leads to the closure
relations
$$
\prod_{y'=1}^{L_2} a(1, y^\prime)^{n_{yy'}} = I, \quad \hbox{for all $1 \leq
y \leq L_2$}.
\eqno (4.5)
$$
The Eqns (4.5) give a presentation of $G$, the structure of which can be
determined from the normal form decomposition of the $L_2 \times L_2$ matrix
$n_{yy'}$. This is considerably easier to handle than the normal form
decomposition of the much larger matrix $\Delta$ needed for an arbitrary ASM.

Even though this is a real computational improvement, the actual calculation,
for arbitrary $L_1$, of the rank of $G$ is nontrivial. Even in the simplest
case $L_2=2$, it depends in a complicated way on the number--theoretic
properties of $L_1$. Details of this case are given in Appendix A.

As to the square lattice where $L_1=L_2=L$, using the above algorithm
we find the following structures of $G$ for the first values of $L$
$$
\eqalignno{
L = 2 \;\;:\;\; G & = Z_{24} \times Z_8, & (4.6a) \cr
L = 3 \;\;:\;\; G & = Z_{224} \times Z_{112} \times Z_4, & (4.6b) \cr
L = 4 \;\;:\;\; G & = Z_{6600} \times Z_{1320} \times Z_8 \times Z_8, &
(4.6c)\cr
L = 5 \;\;:\;\; G & = Z_{102960} \times Z_{102960} \times Z_{48} \times
Z_{16} \times Z_4. & (4.6d) }
$$

This suggests, and it is in fact not difficult to prove, that for an
$L \times L$ square
$$
g = L, \quad {\rm for} \quad L_1 = L_2 = L.
\eqno (4.7)
$$

The idea of the proof of (4.7) is to use the fact that in this case, the
matrix $\Delta$ has exactly $L$ linearly independent eigenvectors
of eigenvalue 4, of which all components can be chosen to be integers. We use
them to construct $L$ independent operators (none can be expressed as
product of powers of the others) $U_k$, $0 \leq k \leq L-1$, such that
$$
U^4_k = I.
\eqno (4.8)
$$
This implies that the number of generators is at least equal to $L$.  Combined
with the inequality (4.3), we get the result.

To write the operators $U_k$ explicitly, we consider the eigenvectors
$n_k(x,y)$ of $\Delta$ of eigenvalue 4
$$
\sum_{(x',y')} \Delta(x,y;x',y') n_k(x',y') = 4n_k(x,y).
\eqno (4.9)
$$
There are $L$ independent solutions to Eq. (4.9), and a possible choice
is to set ($1 \leq x,y \leq L$, $0 \leq k \leq L-1$)
$$
n_k(x,y) = (-1)^x \big[\delta (y-x-k) - \delta(y+x-k) - \delta(x+y -
2L-2+k) + \delta(y-x+k)\big].
\eqno(4.10)
$$
It is then easy to verify by using (4.9) and the relations (2.3) satisfied
by the $a(x,y)$, that the operators $U_k$ defined by
$$
U_k = \prod_{x,y} a(x,y)^{n_k(x,y)},
\eqno (4.11)
$$
all satisfy $U_k^4 = I$. Note that a strong form of
independence of the eigenvectors $n_k(x,y)$ must hold for the
operators $U_k$ to be multiplicatively independent: the eigenvectors must be
linearly independent modulo the lattice $\{\sum_{x',y'} \, m_{x',y'}
\Delta(x,y;x',y') \;:\; m_{x',y'} \hbox{ integers}\}$, on account of the
relations (2.3). The vectors (4.10) are easily seen to satisfy this condition.
This completes the proof.

When $L_1 \not= L_2$, the above construction is easily generalized. If $f =
{\rm gcd}[(L_1+1),(L_2+1)]$, it is easy to check that the matrix $\Delta$
has $f - 1$ independent eigenvectors of eigenvalue 4.
Thus we obtain in this case that the rank $g_{L_1 \times L_2}$ satisfies the
two inequalities
$$
f - 1 \leq g_{L_1 \times L_2} \leq L_2\,.
\eqno (4.12)
$$

That the eigenvectors of integer eigenvalue play a particular role in the
above proof should be clear. An arbitrary eigenvalue is generically an
algebraic number, and the corresponding eigenvector has components which
are also algebraic numbers, so that (4.11) becomes meaningless. However
we show in the next two sections that one can construct toppling invariants
from the spectrum of $\Delta$, which yield the proper setting to generalize
the above idea to any eigenvalue. The corresponding construction is applicable
to any ASM which has a diagonalizable toppling matrix.

\bigskip
\bigskip

\noindent {\bf 5. A reminder of Galois theory.}

In this section, we recall the basic ideas of Galois theory, which can
be used in any ASM, whatever its type. For illustrating the technique,
we shall consider the familiar ASM defined on an $L \times L$ square
lattice. The toppling matrix $\Delta$ is the discrete Laplacian, and
its eigenvalues are easily determined and read
$$
\lambda_{m,n} = 4 - 2\cos {2\pi m \over N} - 2\cos {2\pi n \over N}\,,
\quad 1 \leq m,n \leq L,
\eqno (5.1)
$$
where we have defined
$$
N = 2(L+1).
\eqno (5.2)
$$
The order of the group $G$ is given by
$$
|G| = \prod^L_{m=1} \prod^L_{n=1} \lambda_{m,n}.
\eqno (5.3)
$$
We define the transformations
$$
\sigma_s(\lambda_{m,n}) = \lambda_{m',n'}, \quad \hbox{$s$ coprime with $N$},
\eqno(5.4)
$$
where $m', n'$ are integers, $1 \leq m',n' \leq L$, such that
$\cos (2\pi m'/N) = \cos (2\pi s m/N)$ and $\cos (2\pi n'/N)  = \cos
(2\pi s n/N)$. We see that these transformations act on the set of
eigenvalues by permutation. Moreover, they form a group, and as
$$
\sigma_s \circ \sigma_{s'} = \sigma_{ss'} =\sigma_{s'} \circ \sigma_s \,,
\eqno(5.5)
$$
the group is commutative and isomorphic to the multiplication
modulo $N$ of the numbers coprime with $N$. Finally from the explicit
expression of $\lambda_{m,n}$, the actions of $\sigma_s$ and $\sigma_{-s}$
are the same. Thus the group of all $\sigma_s$ is isomorphic to the group
$Z^*_N$ of invertible integers modulo $N$ (those which are coprime with $N$)
quotiented by the subgroup $\{+1,-1\}$
$$
{\rm Gal}_L = \{\sigma_s \;:\; 1 \leq s \leq L \hbox{ and $s$ coprime
with $N$}\}
\simeq Z^*_N/\{\pm 1\},
\eqno(5.6)
$$
and is of order $\half \varphi(N)$. We call the group (5.6) the Galois
group of this ASM.

It is instructive to group the set of eigenvalues $\lambda_{m,n}$ into
orbits under the Galois group (5.6).
As an example, let us take $L=3$ or $N=8$. The Galois group consists of the
two transformations $s=1$ (the identity transformation) and $s=3$. We find
$\sigma_3(\lambda_{1,1}) = \lambda_{3,3}$ and $\sigma_3(\lambda_{3,3}) =
\lambda_{1,1}$, so that $\{\lambda_{1,1},\lambda_{3,3}\}$ form an orbit
under the Galois group. Computing the product of these two eigenvalues, we
find $\lambda_{1,1}\lambda_{3,3} = 8$. Likewise $\sigma_3(\lambda_{2,2}) =
\lambda_{2,2}$ shows that $\lambda_{2,2}=4$ is an orbit on its own. (Note
that because of the existence of some degeneracies in the eigenvalues,
finding the orbit of $\lambda_{m,n}$ is not merely finding the orbit of
$(m,n)$ under a diagonal multiplication by all $s$. In the example at hand, to
say that $\{\lambda_{1,1},\lambda_{3,3}\}$ is the orbit of $\lambda_{1,1}$
supposes that we have checked that the two eigenvalues are different
complex numbers.) Doing
the same calculation for the other eigenvalues, we can rearrange the product
(5.3) giving the order of $G$ as a product over the orbits (6 in this
case) to find
$$
\eqalign{
L=3 \;: \qquad|G| &= [\lambda_{11} \lambda_{33}]\,[\lambda_{22}]\,
[\lambda_{12} \lambda_{32}]\,[\lambda_{21} \lambda_{23}]\,[\lambda_{13}]\,
[\lambda_{31}]
\cr
&= 8 \cdot 4 \cdot 14 \cdot 14 \cdot 4 \cdot 4 = 100\,352.}
\eqno (5.7)
$$
We can do the same rearrangement for any value of $L$, writing $|G|$ as
a product over orbits under the Galois group,
$$|G| = \prod_{{\rm orbits}\;{\cal O}} \;\big[ \prod_{\lambda_{m,n} \in
{\cal O}} \lambda_{m,n} \big],
\eqno(5.8)
$$
and then each sub--product in square brackets is an integer. This
follows from the fact that each square bracket is by construction
invariant under the whole of the Galois group, in which case the
Galois theorem then states it must be a rational number. Because of
the special form of the numbers $\lambda_{m,n}$ (they are algebraic
integers), all square brackets must even be integers. Not only is each
square bracket equal to an integer, but the Galois theorem asserts
that none of them can be further split: the decomposition (5.8) is the
maximal factorization of $|G|$ as a product of integers with the
property that each integer is the product of a certain number of
eigenvalues.

Let us now put the above considerations in a more general perspective.
A clear account of Galois theory can be found, for example, in [20].

Let $P(x)$ be a finite degree polynomial with integer coefficients, and
let us define the algebraic extension $F[P]$ by adjoining to the field
of rational numbers $Q$
all the roots of $P(x)$. (This kind of algebraic extensions are technically
known as separable normal extensions.)
Thus, $F[P]$ is constructed in two steps: first
we form the set of all rational linear combinations of the roots of $P$,
and second, we promote this set to a field by adding whatever is
needed to make it
closed under multiplication. Hence $F[P]$ contains all roots of $P$ and
all products of roots, but it is always possible to choose a finite
number of base elements (linearly independent over $Q$) of which $F[P]$ is
the linear $Q$--span. The number of base elements is called the degree of the
extension $F[P]$, which has in general no relation with the degree
of $P$. For instance the degree of $F[x^4-2]$ is 8, while
that of $F[x^4-1]$ is 2.

Clearly, $Q$ is a subfield of $F[P]$, and it is thus a well--defined
problem to look for automorphisms of $F[P]$ which leave invariant every
element of $Q \subset F[P]$. The group of all such automorphisms is
called the Galois group of $F[P]$ with respect to $Q$, and noted
Gal$(F[P]/Q)$. The effect of the Galois group on
$F[P]$ is to permute the roots of $P$, although not all permutations
are allowed since they must be automorphisms. An equivalent definition
of Gal$(F[P]/Q)$ is the set of permutations of the roots of $P(x)$ which
preserve all algebraic relations among them. It can be shown that the order
of the Galois group is equal to the degree of $F[P]$. We can now formulate
the Galois Correspondence Theorem: there is a one--to--one correspondence
between the subfields of $F[P]$ and the subgroups of Gal$(F[P]/Q)$. The
bijection goes by associating a subfield $M \subset F[P]$ with the
maximal subgroup $H \subset {\rm Gal}(F[P]/Q)$ which leaves the elements
of $M$ invariant. Hence the bigger $H$, the smaller $M$. For instance
$M=Q$ is associated with $H={\rm Gal}(F[P]/Q)$ by definition of the
Galois group (hence an element of $F[P]$ is in $Q$ iff it is invariant
under the whole Galois group), and
$M=F[P]$ is associated with the trivial subgroup  $H=\{1\}$.

Before closing this digression, we make a final comment about algebraic
numbers and algebraic integers. All the elements of an algebraic
extension of $Q$ are algebraic numbers, which means that they all
satisfy polynomial equations with integer coefficients. However, some of
them may as well satisfy a polynomial equation with integer coefficients
and with coefficient of the highest power equal to 1. These elements
are distinguished in the extension and are called algebraic integers.
For instance, in $F[x^2-2]=Q[\sqrt{2}]$, the number $\sqrt{2}$ is an
algebraic integer whereas $1/\sqrt{2}$ is not. Also the eigenvalues of
a matrix with integer entries are all algebraic integers.
While the extension itself is a field, the
subset of its algebraic integers is only a ring. In $Q$, this amounts to the
usual distinction between the rationals and the integers. An immediate
consequence is that an algebraic integer of $F[P]$ which is invariant under
the whole Galois group is actually an algebraic integer of $Q$, {\it i.e.}
an integer of $Z$.

Let us now see how these general ideas work if we take the extension
$F[{\rm det}(x-\Delta)]$ obtained by adjoining to $Q$ the roots of the
characteristic polynomial of the toppling matrix of an ASM. We choose
the ASM defined in section 3 for an $L \times L$ lattice, with $\Delta$
the discrete Laplacian, but clearly the same analysis can be done in
any model.

We thus consider the algebraic extension $F_L \equiv F[{\rm det}
(x-\Delta_L)]$ obtained by adding to $Q$ the roots $\lambda_{m,n}$ given in
(5.1),
$$
\lambda_{m,n} = 4 - 2\cos {2\pi m \over N} - 2\cos {2\pi n \over N}\,,
\quad 1 \leq m,n \leq L.
\eqno (5.9)
$$
Since $F_L$ contains all rational combinations of the $\lambda_{m,n}$,
it contains $\cos{2\pi m \over N}$ for all $1 \leq m \leq L$ since
$$
\cos{2\pi m \over N} = 2 - {1 \over 2L} \sum_{n=1}^L \lambda_{m,n}.
\eqno(5.10)
$$
We also see from
$$
\cos{2\pi m \over N} \cdot \cos{2\pi n \over N} = {1 \over 2}
\cos{2\pi (m+n) \over N} + {1 \over 2} \cos{2\pi (m-n) \over N},
\eqno(5.11)
$$
that any product of $\lambda_{m,n}$ can be expressed as a rational
combination of 1 and the numbers (5.10), which shows that any element of
$F_L$ can be written as a rational combination of
1, $\cos{2\pi \over N}$, $\cos{4\pi \over N}$, $\ldots$,
$\cos{2\pi L\over N}$. However, this writing is not unique as these $L+1$
cosines are not independent over $Q$, but satisfy
$$
\sum_{i=0}^{p-1} \cos{2\pi (ap^k + bN/p^k + iN/p) \over N} = 0, \quad
0 \leq a \leq {N \over p^k}-1, \; 0 \leq b \leq p^{k-1}-1,
\eqno(5.12)
$$
for every prime power $p^k$ entering the prime decomposition of $N$. Using
these relations, it can be shown that only $\half \varphi(N)$ among the $L+1$
cosines are independent over $Q$. Thus the extension $F_L$, usually denoted
by $Q(\cos{2\pi \over N})$, has degree $\half\varphi(N)$.
A basis of $F_L$ over $Q$ can be taken to be
$\{\cos {2\pi m \over N} \;:\; 0 \leq m \leq \half\varphi(N)-1 \}$.

To find the Galois group of $F_L$, we first note the following identities
$$
\cos{2\pi m \over N} = T_m(\cos{2\pi \over N}),
\eqno(5.13)
$$
where $T_m$ is the $m$--th Chebyshev polynomial of the first kind.
Equation (5.13) shows that the Galois transformation of $\cos{2\pi m \over N}$,
$m \geq 2$, is determined by that of $\cos{2\pi \over N}$. The next
step is to notice that, since the Galois group acts on the roots
$\lambda_{m,n}$ by permutations, the Galois transformations of
$\cos{2\pi \over N}$ must be in the set of $\cos{2\pi s \over N}$,
$1 \leq s \leq L$. Hence suppose that $\sigma_s(\cos{2\pi \over N}) =
\cos{2\pi s \over N}$ is a Galois transformation for $s$ between 1 and $L$.
Being a group element, $\sigma_s$ must have an inverse which we denote by
$\sigma_{s'}=\sigma_s^{-1}$. It satisfies $\sigma_{s'}(\cos{2\pi s\over N})
= \cos{2\pi ss' \over N} = \cos{2\pi \over N}$, hence $ss'=\pm 1 \bmod N$.
This is impossible if $s$ and $N$ have a common factor, while if $s$ is
coprime with $N$, there is a unique $s'$ between 1 and $L$ satisfying
either $s'=s^{-1} \bmod N$ or $s'=-s^{-1} \bmod N$. Thus
the Galois group contains at least the transformations $\sigma_s$ for $s$
between 1 and $L$ and coprime with $N$. The number of such $s$ is equal
to $\half \varphi(N)$, and since this must also be the order of the Galois
group (equal to the degree of $F_L$), one recovers the symmetry
group of (5.6)
$$
{\rm Gal}_L = \{\sigma_s \;:\; 1 \leq s \leq L \hbox{ and $(s,N)=1$}\}.
\eqno(5.14)
$$

We have completely identified the extension $F_L$ and determined its
Galois group. The last point concerns the ring of algebraic integers
of $F_L$. This is a far more difficult question, and we just quote the
result: the algebraic number $w=a_0+\sum_{m=1}^{\half \varphi(N)-1} 2a_m
\cos{2\pi mJ\over N}$ of $F_L$ is an algebraic integer if and only if
all coefficients $a_m$ are integers of $Z$. This ring of integers is
usually denoted by $Z(\cos{2\pi \over N})$. In
particular, as noted above, the eigenvalues $\lambda_{m,n}$ are all
algebraic integers.

Looking back at the decomposition (5.8) of $|G|$ as a product of orbits
of eigenvalues under the Galois group, all the properties we mentioned
easily follow from the above facts. The product of eigenvalues belonging
to an orbit is by definition Galois invariant and thus must be an
rational number. Since the eigenvalues are algebraic integers, this
rational number must in fact be an integer. Finally the decomposition
is maximal by definition of the orbits as the smallest subsets invariant
under the Galois group.

Similarly to the decomposition of $G={\rm Det}\,\Delta$, we may consider the
factorization of the characteristic polynomial of $\Delta$ in factors which
are all invariant under the Galois group
$$
P_L (x) = \prod^L_{m,n=1} (x - \lambda_{m,n}) =
\prod_{{\rm orbits}\;{\cal O}} \;\big[ \prod_{\lambda_{m,n} \in
{\cal O}} (x - \lambda_{m,n}) \big].
\eqno (5.15)
$$
In this case, the Galois theory states that the polynomial within each
square bracket, call it $P_{\cal O}(x)$, has integer coefficients, and
that the decomposition (5.15) is maximal, namely one cannot perform a
further splitting into polynomials with rational coefficients (or in
other words, all $P_{\cal O}(x)$ are irreducible over the field of
rationals).

Examples of decomposition (5.15) for small values of $L$ are
$$
\eqalignno{
&P_2 (4-x) = x^2 (x^2 - 4)\,, & (5.16a) \cr
&P_3 (4-x) = -x^3 (x^2 - 2)^2 (x^2 - 8)\,, & (5.16b) \cr
&P_4 (4-x) = x^4 (x-1)^2 (x+1)^2 (x^2-5)^2 (x^2-2x-4) (x^2+2x-4)\,,
& (5.16c) \cr
& \;\;\;\;\;\; \vdots & \cr
&P_7 (4-x) = -x^7 (x^2-2)^2 (x^2-8) (x^4-4x^2+2)^2 (x^4-8x^2+8)^2 \cr
& (x^4-16x^2+32) (x^4-8x^2-8x-2)^2 (x^4-8x^2+8x-2)^2. & (5.16d)}
$$
The number $F(L)$ of factors in the decompositions (5.8) or (5.15) can be
explicitely computed. The behaviour of $F(L)$ is rather chaotic, as
can be checked from Table I, where we give some of its values.
It strongly depends on the prime decomposition of
$(L+1)$, and for large $L$, $F(L)$ increases linearly with $L$.
An analytic formula for $F(L)$ is given in Appendix B.

\bigskip
$$
\vbox{\offinterlineskip
\halign{&\vrule#&\strut\ #\ \cr
\multispan{17}\hfil{\bf TABLE I}: Number $F(L)$ of irreducible \hfil\cr
\multispan{17}\hfil factors in det$(x-\Delta)$. \hfil\cr
\noalign{\medskip}
\noalign{\hrule}
height3pt&\omit&&\omit&&\omit&&\omit&&\omit&&\omit&&\omit&&\omit&\cr
&\hfil $L$\hfil&&\hfil $F(L)$\hfil&&\hfil $L$\hfil&&\hfil $F(L)$
\hfil&&\hfil $L$\hfil&&\hfil $F(L)$\hfil&&\hfil $L$\hfil&&\hfil $F(L)$
\hfil&\cr
height3pt&\omit&&\omit&&\omit&&\omit&&\omit&&\omit&&\omit&&\omit&\cr
\noalign{\hrule}
height3pt&\omit&&\omit&&\omit&&\omit&&\omit&&\omit&&\omit&&\omit&\cr
&\hfil 1\hfil&&\hfil 1\hfil&&\hfil 6\hfil&&\hfil 16\hfil&&\hfil
21\hfil&&\hfil 74\hfil&&\hfil 26\hfil&&\hfil 112\hfil&\cr
height3pt&\omit&&\omit&&\omit&&\omit&&\omit&&\omit&&\omit&&\omit&\cr
&\hfil 2\hfil&&\hfil 4\hfil&&\hfil 7\hfil&&\hfil 19\hfil&&\hfil
22\hfil&&\hfil 64\hfil&&\hfil 27\hfil&&\hfil 112\hfil&\cr
height3pt&\omit&&\omit&&\omit&&\omit&&\omit&&\omit&&\omit&&\omit&\cr
&\hfil 3\hfil&&\hfil 6\hfil&&\hfil 8\hfil&&\hfil 28\hfil&&\hfil
23\hfil&&\hfil 122\hfil&&\hfil 28\hfil&&\hfil 84\hfil&\cr
height3pt&\omit&&\omit&&\omit&&\omit&&\omit&&\omit&&\omit&&\omit&\cr
&\hfil 4\hfil&&\hfil 12\hfil&&\hfil 9\hfil&&\hfil 34\hfil&&\hfil
24\hfil&&\hfil 88\hfil&&\hfil 29\hfil&&\hfil 192\hfil&\cr
height3pt&\omit&&\omit&&\omit&&\omit&&\omit&&\omit&&\omit&&\omit&\cr
&\hfil 5\hfil&&\hfil 18\hfil&&\hfil 10\hfil&&\hfil 28\hfil&&\hfil
25\hfil&&\hfil 90\hfil&&\hfil 30\hfil&&\hfil 88\hfil&\cr
height3pt&\omit&&\omit&&\omit&&\omit&&\omit&&\omit&&\omit&&\omit&\cr
\noalign{\hrule}
}}
$$

\bigskip
Finally, let us note that the Galois group (5.14) also acts on the $L
\times L$ lattice on which the ASM is defined. This action is defined by
$\sigma_s(x,y) = (x',y')$ where $\cos {2\pi x' \over N} = \cos {2\pi
sx \over N}$ and $\cos {2\pi y' \over N} = \cos {2\pi ys \over N}$.
The operator $\sigma_s$ permutes the lattice sites. In
turn, this induces an action on the set of all configurations of the ASM
by setting
$$
\sigma_s C \;\;: \qquad \sigma_s(z_i) = z_{\sigma_s(i)}.
\eqno(5.17)
$$
So we can also look at the Galois transformations either as automorphisms
of the lattice or as automorphisms of the set of all configurations of the
ASM. In fact, they even define automorphisms of $\cal R$, because,
although $\sigma_s C$ is not necessarily recurrent even if $C$ is, there is
a unique recurrent configuration equivalent to $\sigma_s C$.

\bigskip
\bigskip

\noindent {\bf 6. Toppling invariants from eigenvectors.}

Using the basic results of Galois theory briefly recalled in section 5,
we study here toppling invariants constructed from the (left) eigenvectors of
$\Delta$. This can be done for any ASM with diagonalizable toppling matrix,
but for definiteness, we proceed with the ASM defined in section 4, for
a square $L \times L$ lattice.

Thus $\Delta$ is the discrete Laplacian, whose eigenvalues $\lambda_{m,n}$
were given in (5.1), and whose eigenvectors read
$$
v_{m,n}(x,y) = {\sin{2\pi mx \over N} \over \sin{2\pi m \over N}} \cdot
{\sin{2\pi ny \over N} \over \sin{2\pi n \over N}}, \qquad 1 \leq m,n
\leq L.
\eqno(6.1)
$$
The normalization is purely conventional, and will be explained below.

For each eigenvector $v_{m,n}$, we define an algebraic toppling invariant by
taking its inner product with the $z$ vector of an arbitrary stable
configuration, and by evaluating the result modulo the eigenvalue
$\lambda_{m,n}$ of $v_{m,n}$
$$
A_{m,n}(C) = \sum_{x,y} \, v_{m,n}(x,y) \, z_{x,y} \; \bmod \lambda_{m,n}.
\eqno(6.2)
$$
Since the eigenvalues and the eigenvector components are in general algebraic
numbers in the algebraic extension $F[{\rm det}(x-\Delta)]$, equal to
$Q(\cos{2\pi \over N})$ in this case, the
congruence (6.2) must be understood in that extension. Moreover, modulo
$\lambda_{m,n}$ means modulo all multiples $w \lambda_{m,n}$ with
$w$ in the ring of algebraic integers of $Q(\cos{2\pi \over N})$.

Clearly under the toppling at site $(x_0,y_0)$, we have
$A_{m,n} \longrightarrow A_{m,n} + \lambda_{m,n} v_{m,n}(x_0,y_0)$,
so that $A_{m,n}(C)$ is invariant provided
$v_{m,n}(x_0,y_0)$ is an algebraic integer for all $x_0,y_0$. This was
precisely the reason for choosing the normalization (6.1). On account
of the fact that, for integer $x$, $\sin{\alpha x} \over \sin{\alpha}$
is a polynomial of $2 \cos{\alpha}$ with integer coefficients, the
components of $v_{m,n}$ are indeed algebraic integers of
$Q(\cos{2\pi \over N})$ (see section 5).

Thus the invariants $A_{m,n}$ are all valued in $Z(\cos{2\pi \over N})$
where the congruence (6.2) is to be taken. To give an explicit example,
consider $L=3$, $N=8$. The relevant extension is $Q(\cos{2\pi \over 8})
=Q(\sqrt{2})$. Choosing for instance $m=n=1$, one obtains $\lambda_{1,1}=
4-2\sqrt{2}$, $v_{1,1}(x,y) = (1,\sqrt{2},1) \otimes (1,\sqrt{2},1)$,
so that
$$\eqalign{
A_{1,1}(C) = z_{1,1} + \sqrt{2}z_{1,2} + z_{1,3} + \sqrt{2}z_{2,1} +
2&z_{2,2} + \sqrt{2}z_{2,3} + z_{3,1} \cr
&+ \sqrt{2}z_{3,2} + z_{3,3} \quad \bmod 4-2\sqrt{2}. \cr}
\eqno(6.3)
$$
It can be seen that this invariant takes 8 different values by setting
$z_{1,1}=k$ for $k=0,1,\ldots,7$, and all other $z_{x,y}=0$. Indeed $8=
(4+2\sqrt{2})(4-2\sqrt{2})$ is the smallest positive integer equal to
a multiple of $4-2\sqrt{2}$. Also recall from section 5 that
$\lambda_{1,1}$ belongs to the orbit ${\cal O}=\{\lambda_{1,1},
\lambda_{3,3}\}$
under the Galois group, and that the associated irreducible polynomial
is $P_{\cal O}(x) = (x-\lambda_{1,1})(x-\lambda_{3,3}) = x^2-8x+8$.
So we have in this case that $P_{\cal O}(0) = 8$ is the number of values
taken by the invariant $A_{1,1}$.

It seems that we have defined $L^2$ invariants $A_{m,n}$, one for each
eigenvalue. However some of them are related by Galois transformations,
and are thus not independent. Indeed if $\lambda_{m,n}$ and
$\lambda_{m',n'}$ are in the same Galois orbit, the corresponding
eigenvectors are related by a Galois transformation $\sigma$, and so are the
corresponding invariants, $A_{m',n'}=\sigma(A_{m,n})$. So in fact we associate
an invariant not to each eigenvalue, but rather to each orbit $\cal O$ under
the
Galois group, or equivalently to each irreducible factor $P_{\cal O}(x)$ of the
characteristic polynomial of $\Delta$. We thus have $F(L)$ (the number of
orbits) algebraic invariants. Moreover, because of the normalization we chose
in
(6.1) for the eigenvectors,  each invariant is linear in the height of the left
top corner ($x=y=1$) with coefficient 1:
$$
A_{m,n}(C) = z_{1,1} + \ldots \; \bmod \lambda_{m,n}.
\eqno(6.4)
$$
Since 1 is not an eigenvalue, we see that none of the $A_{m,n}$ is trivial.
However one cannot say in general the number of values they take.
If one evaluates $A_{m,n}(C)$ at those configurations for which $z_{x,y}=0$
for all $(x,y) \neq (1,1)$ and $z_{1,1}=k \in N$, one sees that $A_{m,n}$
takes positive integer values, but because of the congruence, different
$k$'s will give $A_{m,n}$ the same value. Denoting by $k_{\rm max}(m,n)$
the smallest positive integer equal to zero modulo $\lambda_{m,n}$, $A_{m,n}$
will take exactly $k_{\rm max}(m,n)$ different values. To determine
that integer is not easy. Clearly an upper bound is given by the product of
$\lambda_{m,n}$ with all its Galois conjugates, an integer by construction.
This gives $k_{\rm max}(m,n) \leq  |P_{\cal O}(0)|$ where $\cal O$
is the orbit containing $\lambda_{m,n}$. It can be shown that the equality
holds whenever $\lambda_{m,n}$ is not the product of two integers belonging
to two different subrings of $Z(\cos{2\pi \over N})$. In the other cases,
the exact value of $k_{\rm max}(m,n)$ depends on the factorization
properties of $\lambda_{m,n}$. [For instance, for $L=4$, $N=10$, we
have $Q(\cos{2\pi \over 10})=Q(\sqrt{5})$. For $\lambda_{1,1}=5-\sqrt{5}$,
we find $k_{\rm max}=10$, while the associated value of $P_{\cal O}(0) =
(5-\sqrt{5})(5+\sqrt{5})$ is equal to 20. The reason is that $5-\sqrt{5}=
2 {5-\sqrt{5} \over 2}$ is the product of 2, an integer of $Z$ and of
${5-\sqrt{5} \over 2}$, an integer of $Z(\sqrt{5})$.]

The next question is whether the invariants $A_{m,n}$ are linearly
independent. This is easily answered negatively. There are $F(L) > L$
invariants and we have shown that none of them is trivial, so that if
they were independent, the group $G$ would be the product of at least
$F(L)$ cyclic groups since it would have to contain the subgroup
$\times_{m,n} \, Z_{k_{\rm max}(m,n)}$. (Remember from section 3 that
toppling invariants provide additive representations of $G$.) This is
of course impossible since, from section 4, we know that $G$ is the product
of exactly $L$ cyclic factors.

Even if the set of $A_{m,n}$ is not linearly independent, we can still
consider the subset of independent ones, hopefully of cardinality equal to
$L$, and see if those provide a complete labelling
of the set of recurrent configurations. This too can be easily shown to
fail. If they did, the total number of different values they take should be
equal to Det$\,\Delta$. But since Det$\,\Delta = \prod_{\cal O} |P_{\cal O}
(0)|$, the right number of values is reached only if we take all the
invariants and if, for each of them, $k_{\rm max}(m,n) = |P_{\cal O}(0)|$.
This is clearly impossible from the linear dependence among some of them.

Thus the invariants $A_{m,n}$ are neither independent nor complete. In view
of the results of section 3, this is hardly surprising since the Smith
normal form is not related to a spectral decomposition.
A natural question is then whether one can find a complete set of
algebraic invariants, analogous to the complete set of integer invariants
$I_i$ defined in (3.11) in terms of the Smith normal form. It is indeed
possible by using the following algorithm.

We note that since the invariants involve congruences, there is no need
to use the exact eigenvectors. We can define invariants by setting
$$
A_\lambda (C) = \sum_{x,y} \, v_\lambda (x,y) \, z_{x,y} \; \bmod \lambda ,
\eqno(6.5)
$$
where we now take for $v_\lambda$ a vector with the property that
$$
v_\lambda \cdot \Delta = 0 \; \bmod \lambda.
\eqno(6.6)
$$
With no loss of generality, we can assume that the components of $v_\lambda$
are in the ring of integers of some algebraic extension (otherwise
$\lambda$ is simply rescaled), so that we can see the vectors
$v_\lambda$ as left eigenvectors of zero eigenvalue over a finite ring.
We will call them modular eigenvectors (with zero eigenvalue).
We look for a set of such vectors, as small as possible and with as large
values of $\lambda$ as possible, in order to generate a complete set of
invariants with the fewest invariants. To avoid irrelevant overall factors
(if $v_\lambda$ is a modular eigenvector modulo $\lambda$, $2v_\lambda$ is
a modular eigenvector modulo $2\lambda$), we may assume that at least one
component of $v_\lambda$ is coprime with $\lambda$, which we can then
choose equal to 1. The system (6.6) is overdetermined since it yields $L^2$
equations for only $L^2-1$ components of $v_\lambda$ (one of them was
set equal to 1). The extra equation can be seen as a constraint on the
values of $\lambda$. By the Chinese Remainder theorem, it is
sufficient to look for solutions of (6.6) for $\lambda$ a prime power.
For if $v_{\lambda_1}$ and $v_{\lambda_2}$ are modular eigenvectors modulo
$\lambda_1$ and $\lambda_2$ respectively, with $\lambda_1$ and $\lambda_2$
coprime, $v=v_{\lambda_1}\lambda_2 +
v_{\lambda_2}\lambda_1$ is a modular eigenvector modulo $\lambda_1
\lambda_2$. Determining all independent solutions modulo prime powers
and systematically using the Chinese Remainder theorem to obtain the
smallest number of modular eigenvectors, we get a set of eigenvectors
$\{v_{\lambda_i}\}$ where $\lambda_i$ is a multiple of $\lambda_{i+1}$.
((6.6) implies that $\lambda_1$, hence all $\lambda_i$, divides
the determinant of $\Delta$.)
The corresponding set of invariants $\{A_{\lambda_i}\}$ defines a complete
set of toppling invariants.

As an example, consider once more the $2 \times 2$ case. Let us choose
the $v_\lambda$'s and the $\lambda$'s in $Z$, and assume
$$
A_\lambda (C) = z_1 + bz_2 + cz_3 + dz_4 \; \bmod \lambda,
\eqno(6.7)
$$
where the configuration $C$ is represented by $\pmatrix{z_1 & z_2 \cr
z_3 & z_4}$. The equation (6.6) implies $c=4-b \bmod \lambda$ and
$d=4b-1 \bmod \lambda$, as well as
$$
8b-16 = 16b-8 = 0 \bmod \lambda.
\eqno(6.8)
$$
For $\lambda = p^m$ with $p$ an odd prime, (6.8) is equivalent to
$b = 2 = 2^{-1} \bmod p^m$, which has no solution unless $p^m=3$, in which
case one finds one invariant,
$$
A_3 (C) = z_1 + 2z_2 + 2z_3 + z_4 \bmod 3.
\eqno(6.9)
$$
For
$\lambda = 2^m$, the two equations (6.8) are trivially satisfied for $2^m=8$,
but cannot be satisfied for $2^m \geq 16$ ($16b-8=-8 \neq 0 \bmod 16$). Thus
the maximal value of $\lambda$ is 8, and one finds two independent
invariants modulo 8 ($b=1,-2$)
$$\eqalignno{
A_8(C) &= z_1 + z_2 + 3z_3 + 3z_4 \bmod 8. &(6.10a) \cr
A_8'(C) &= z_1 - 2z_2 - 2z_3 - z_4 \bmod 8, &(6.10b) \cr}
$$
Using the Chinese Remainder theorem, we obtain two invariants, modulo
24 and 8 respectively:
$$\eqalignno{
A_{24}(C) &= 8A_3(C) + 3A_8'(C) = 11z_1 + 10z_2 + 10z_3 + 5z_4 \bmod 24,
&(6.11a) \cr
A_8(C) &= z_1 + z_2 + 3z_3 + 3z_4 \bmod 8. &(6.11b) \cr}
$$
As expected, one recovers the invariants (3.8) constructed out of
the Smith normal form of $\Delta$, $I_1=5A_{24}$ and $I_2=3A_8$.
Clearly, this is a general fact: choosing to work on $Z$, the above
algorithm allows to compute the elementary divisors of $\Delta$ (the
$\lambda_i$'s), as well as the relevant lines of the matrix $A^{-1}$
needed to obtain the invariants (see (3.11)), since from (3.9) and (3.10),
the rows of $A^{-1}$ are precisely modular eigenvectors modulo the
elementary divisors of $\Delta$.
However the algorithm is completely general regarding the ring of integers we
want to work in. In particular, one can find a complete set of invariants with
values in the ring of integers of any algebraic extension, something we
will not pursue here since the algebraic invariants we would so obtain can be
read off from the $Z$--valued ones (by decomposing the elementary divisors
$d_i \in Z$ into a product of powers of prime ideals of the extension).

As a conclusion to this section, we found it rather attractive to use
Galois theory in order to associate toppling invariants with irreducible
factors of the characteristic polynomial of $\Delta$, to the end of
extracting information about $G$ from the Galois decomposition of this
polynomial. Although this method can be useful, the usefulness being
dependent on the spectrum of $\Delta$, it generally provides
only partial information on $G$.

\bigskip
\bigskip

\noindent {\bf 7. Some related problems}

In this section, we comment briefly on two intriguing questions related
to sandpile automata on finite square lattices.

\bigskip \noindent
QUESTION 1.  Wiesenfeld {\it et al} [15] have studied a deterministic
version of the sandpile automaton considered here, in which the particle
addition is always done at the central site of a $(2L+1) \times (2L+1)$
square lattice. They
observed, and it is easy to prove, that starting from an initial arbitrary
configuration, once the transient configurations are gone, the system
shows a cyclic behavior, and the period of the cycle $T_L$ is independent
of the initial configuration. The exact dependence of $T_L$ on $L$
is still quite puzzling.

Wiesenfeld {\it et al} have numerically determined $T_L = 4,16,104, 544,
146\,248, 7\,889\,840$ for $L=0$ to 5, and found that $T_L$ increases at
a much slower rate than Det$\,\Delta$, $T_L \sim \exp{0.44\,L^2}$ against
Det$\,\Delta \sim \exp{4.67 \,L^2}$. The
period for $L$ up to 9 has been determined by Markosova and Markos [16].
It is easy to see that $T_L$ is the order of the operator $a(L+1,L+1)$ on
the space $\cal R$ of recurrent configurations
and so is the smallest positive integer such that
$$
a(L+1,L+1)^{T_L} = I.
\eqno (7.1)
$$
If $a(L+1,L+1)^{T_L}=I$ on $\cal R$, we obtain that all the toppling
invariants (3.3) must be equal to zero modulo 1 at the configuration which is
zero everywhere except at the central site, where it is $T_L$. Multiplying
the invariants by Det$\,\Delta$ and defining the integer matrix
$E=({\rm Det}\, \Delta)\Delta^{-1}$, we obtain that $T_L$ is the smallest
positive integer satisfying
$$
E_{i,(L+1,L+1)}\,T_L = 0 \; \bmod ({\rm Det}\,\Delta),\quad\hbox{for all $i$}.
\eqno(7.2)
$$
Setting $M = {\gcd}\,\{E_{i,(L+1,L+1)}\,:\,1 \leq i \leq 2L+1 \}$,
it is easy to see that
$$
T_L = ({\rm Det}\,\Delta)/M.
\eqno (7.3)
$$

Since $T_L$ is independent of the initial
configuration, we may choose a configuration which has the symmetry of the
square. This symmetry is preserved
under symmetrical topplings, and under addition of sand at the central site,
hence we  can restrict ourselves
to ${\cal R}_{\rm sym}$, the subspace of recurrent configurations having the
symmetry of a square.  On a $(2L+1) \times (2L+1)$ lattice, a symmetrical
configuration is completely specified by the heights in an octant
having ${(L+1) (L+2) \over 2}$ sites. Instead of studying the symmetrical
configurations on a square lattice, we can just as well study all recurrent
configurations on an octant $O_L$ with a new toppling matrix
$\Delta_{\rm sym}$ given by
$$
(\Delta_{\rm sym})_{ij} = \sum_{j'} \Delta_{i,j'}, \quad {\rm for}\,
i,j \in O_L,
\eqno (7.4)
$$
where the sum over $j'$ is over sites which are related to $j$ by the
symmetries of the square (dihedral group of order 8). Even though it does not
satisfy $\sum_i (\Delta_{\rm sym})_{ij} \geq 0$ for all $i$, it can be shown
that the general results of [6] hold for this case, so that for example
$$
|{\cal R}_{{\rm sym}}| = {\rm Det}\,\Delta_{\rm sym} =  |G_{\rm sym}|,
\eqno (7.5)
$$
where $G_{\rm sym}$ is the abelian group generated by the operators $a_i\;, i
\in O_L$, subjected to the relations (2.3) with $\Delta$ replaced by
$\Delta_{\rm sym}$. A simple calculation shows that
$$
{\rm Det}\,\Delta_{\rm sym} = \prod_{m \leq n} \lambda_{m,n},
\eqno (7.6)
$$
where the product over $m$ and $n$ is over {\it odd} values of
$m \leq n$ between 1 and $(2L+1)$, and with $\lambda_{m,n}$ given in (5.1). For
large $L$, $|G_{\rm sym}|$ varies as $|G|^{1/8} \sim \exp{0.58\,L^2}$, so that
$T_L$ increases at a substantially lower rate than $|G_{\rm sym}|$. As in
section 3, the structure of $G_{\rm sym}$ can be determined by computing the
elementary divisors of $\Delta_{\rm sym}$. We find for the first four values
of $L$ $$
\eqalignno{
& L = 1 \;\;:\;\; G_{\rm sym} = Z_{16} \times Z_2, &(7.7a) \cr
& L = 2 \;\;:\;\; G_{\rm sym} = Z_{104} \times Z_4 \times Z_2, &(7.7b) \cr
& L = 3 \;\;:\;\; G_{\rm sym} = Z_{544} \times Z_{32} \times Z_2 \times Z_2,
&(7.7c) \cr
& L = 4 \;\;:\;\; G_{\rm sym} = Z_{146248} \times Z_8 \times Z_4
\times Z_2 \times Z_2. &(7.7d)}
$$
We see that the order of the largest cyclic group in $G_{\rm sym}$ is
apparently nothing but the period $T_L$. In terms of the invariants (3.11)
constructed from the Smith normal form of $\Delta_{\rm sym}$, this would
mean that the element of the first row of $A^{-1}$ corresponding to the
central site is coprime with the largest elementary divisor $d_1^{\,\rm sym}$,
implying $T_L = d_1^{\,\rm sym}$. What is certainly true is that $T_L$ must
divide $d_1^{\,\rm sym}$ since $a(L+1,L+1)$ generate a cyclic subgroup of
$G_{\rm sym}$, the largest one being of order $d_1^{\,\rm sym}$.
Although there
are plausibility arguments that $T_L$ is actually equal to
$d_1^{\,\rm sym}$, the proof (or disproof) of this simple fact is still
lacking.

The question of the rank of $G_{\rm sym}$ can also be addressed. However here,
an argument based on the eigenvectors of eigenvalue 4, similar to the one
we used in section 4, does not work since the degeneracy of the eigenvalue 4
is only $[{L \over 2}]+1$, with $[x]$ the integral part of $x$. (It
nevertheless implies that so many cyclic factors of $G_{\rm sym}$ have
order multiple of 4.) But on the other hand, it is easy to see that there
are exactly $L+1$ independent modular eigenvectors modulo 2 (of
$\Delta_{\rm sym}$), which shows that the rank of $G_{\rm sym}$ is $L+1$.
($L+1$ is also an upper bound since all operators $a(x,y)$ for $(x,y) \in
O_L$ can be expressed in terms of $a(1,y)$ for $y=1,2,\ldots,L+1$.)
The independent modular eigenvectors have only
one non--zero component, at any of the $L+1$ positions on the principal
diagonal of $O_L$: $v_{\lambda=2}^{(k)}(x,y) = \delta_{x,k}\,\delta_{y,k}$
for $1 \leq k \leq L+1$. This proves that each cyclic subgroup of
$G_{\rm sym}$ has order multiple of 2.

The fact that $T_L$ must be a divisor of $|G_{\rm sym}|$ already gives
us a fairly good upperbound estimate of it. It is instructive to look at the
values of $|G_{\rm sym}|/T_L$ for small $L$. From the results of Markosova
{\it et al} [10], we see that for $L = 0$ up to $8$, these values are
$1,2,2^3,2^7,2^7,2^{11},2^9,2^{22},2^{15}\cdot 2701$. Note that they
show a rather irregular, non--monotonic behavior with $L$. They have a
large power of 2 as divisor, but sometimes have other factors as well.

The fluctuations in $T_L$ appear to be related to the appearance of
degeneracies in the eigenvalue spectrum of $\Delta_{\rm sym}$. For
example, the fact that the eigenvalue 4 is $([{L \over 2}]+1)$--fold
degenerate and the existence of $L+1$ modular eigenvectors modulo 2
imply that $T_L$ is a divisior of $2^{-(L+[L/2])}|G_{\rm sym}|$.
Another accidental degeneracy in the spectrum of $G_{\rm sym}$ occurs when
$2L+2$ is a multiple of 6. Writing $2L+2=6\ell$, we can easily
verify that
$$
\eqalign{
\lambda_{2\ell - m, 2\ell + m} & = \lambda_{3\ell , m} \cr
\lambda_{4\ell - m, 4\ell + m} & = \lambda_{3\ell , 6\ell - m}.}
\eqno (7.8)
$$
In the Galois factorization of $|G_{{\rm sym}}|$ this implies that there are
some factors which are repeated. For example, for $L=8$, the factor 2701
occurs twice in $|G_{{\rm sym}}|$. While for a given value of $L$, the
degeneracies, and the group structure of $G_{{\rm sym}}$ can be explicitly
determined, it seems difficult at this stage to say much more about this
``irregular'' variation of $T_L$ with $L$ for general $L$.

\bigskip \medskip \noindent
QUESTION 2. This concerns the structure of the identity configuration.
Even the $L \times L$ square case shows nontrivial features. The
identity configuration $C_{\rm id}$ is the unique recurrent configuration
which is equivalent under toppling to the configuration with all heights
zero. We have already noted that all toppling invariants $I_i (C_{\rm id})$
must be zero. It also implies that $C_{\rm id}$ is the
unique recurrent configuration such that $T_i \equiv \sum_j (\Delta^{-1})_{ij}
z_j (C_{\rm id})$ are integers for all $i$ (by using the invariants (3.3)).
This integer vector has the interesting
interpretation that $T_i$ is the number of topplings occurring at $i$ when
$C_{\rm id}$ is added to itself.

The identity configuration shows complicated fractal structures. Some
color--coded pictures of these may be found in [21]. We mention here two
remarkable properties of the identity.\footnote{*}{\tenrm We are indebted to
Jean-Louis Ruelle for having run a program to compute the identity, and for
having pointed out these two regularities.}  First, in the central area of a
$2L \times 2L$ lattice, there is a whole square where all sites have maximal
height, equal to 3. The linear size of this central square was numerically
measured for $L$ up to 100, and goes like $4L/5$, so that the central square
covers approximately $4/25$ of the whole lattice.

The second property, even more remarkable, is that the identity configuration
on the $(2L+1)\times (2L+1)$ lattice appears to be related in a very simple way
to  that on the $2L \times 2L$ lattice. Indeed, if we divide the configuration
$C_{\rm id}^{(2L)}$ into four equal squares and pull them apart by one
lattice spacing so as
to leave a cross in the middle, we get $C_{\rm id}^{(2L+1)}$ provided the
heights on the cross are properly assigned. In obvious notation, if
$$
C^{(2L)}_{\rm id} = \pmatrix{B_1 & B_2 \cr B_3 & B_4}, \quad
\hbox{($B_i$ are $L \times L$ blocks),}
\eqno (7.9)
$$
where the four blocks $B_i$ are related by the symmetry transformations
of the square (since $C_{\rm id}$ has that symmetry), then
$$
C_{\rm id}^{(2L+1)} = \pmatrix{B_1 & R_1 & B_2 \cr R_2 & z_{\rm mid} & R_3 \cr
B_3 & R_4 & B_4}, \quad \hbox{($R_i$ are $1 \times L$ rows).}
\eqno (7.10)
$$
In addition the height at the center $z_{\rm mid}$ is always 0, and the
branches of the cross given by the $R_i$ (also related by symmetry
transformations) have a simple structure. The first
instance of this phenomenon occurs when going from the $2 \times 2$ to
the $3 \times 3$ lattices:
$$
C^{(2)}_{\rm id} = \pmatrix{2&2 \cr 2&2} \quad \longrightarrow \quad
C^{(3)}_{\rm id} = \pmatrix{2&1&2 \cr 1&0&1 \cr 2&1&2}.
\eqno(7.11)
$$
More identity configurations are listed in Appendix C, for which this property
may be checked (as well as the first one mentioned above).

In addition, the array $T_i$ defined in a previous paragraph also has the
``cross'' property (7.9--10). For instance,
$$
T^{(6)}=\pmatrix{2&3&4&4&3&2 \cr 3&5&6&6&5&3 \cr 4&6&7&7&6&4 \cr
4&6&7&7&6&4 \cr 3&5&6&6&5&3 \cr 2&3&4&4&3&2}  \quad \longrightarrow \quad
T^{(7)}=\pmatrix{2&3&4&4&4&3&2 \cr 3&5&6&6&6&5&3 \cr 4&6&7&7&7&6&4 \cr
4&6&7&7&7&6&4 \cr 4&6&7&7&7&6&4 \cr 3&5&6&6&6&5&3 \cr 2&3&4&4&4&3&2}.
\eqno(7.12)
$$
Because the $C_{\rm id}$ and $T$ arrays are related by
$C_{\rm id} = \Delta \,T$, the fact that both have the ``cross'' property
fixes their values on the branches of the cross (for odd lattices). For
instance, the rows in $T$ corresponding to the $R_i$ (see (7.10))
must be equal to the rows bordering them, as shown
in (7.12) for $2L+1=7$. The first property mentioned for $C_{\rm id}$,
namely the existence of a big central square with constant values,
does not hold
for $T$. There is in fact in $T$ a central square of constant values, but its
linear size cannot exceed 3 if $C_{\rm id}$ is to be recurrent at all
(see (7.12)).
As a final remark, we observed that the total number of topplings occurring
when the identity is added to itself, {\it i.e.} $\sum_i \,T_i^{(L)}$, equal to
160  and 235 for $L=6$ and 7, grows like a power of $L$, with an exponent
close to 4 ($\sim 3.9$).

Similar geometric structures are found when ASM are allowed to relax from
special unstable states, say all heights equal to 4 [14]. These fractal
structures are not well understood yet.

\vskip 1.5truecm \noindent
The work of S. Sen was supported in part by EOLAS Scientific Research
programme SC/92/206. P. Ruelle acknowledges the Scholarship of
the Dublin Institute for Advanced Studies, where most of his work
was carried out.

\vskip 2.5truecm
\noindent
{\bf{Appendix A}}

\bigskip
In this appendix, we determine the structure of the group $G$
for the ASM defined on the $L \times 2$ lattice.  In this case, the particle
addition operators $a(x,y)$ satisfy the relations
$$
a(x+1, 1) = a^4(x,1) \; a^{-1}(x-1,1) \; a^{-1}(x,2),
\eqno (A1)
$$
and
$$
a(x+1, 2) = a^4(x,2) \; a^{-1}(x-1,2) \; a^{-1}(x,1).
\eqno (A2)
$$
Here $1 \leq x \leq L$, and by convention we assume that
$$
a(0,y) = a(L+1,y) = I, \quad \hbox{for $y=1,2$}.
\eqno (A3)
$$
Using the recursion relations (A1)
and (A2) we can express $a(x,1)$ and $a(x,2)$ as product of powers of
$a \equiv a(1,1)$ and $b \equiv a(1,2)$.  Explicitely we have
$$
\eqalignno{
a(x,1) & = a^{\mu_x} \; b^{-\nu_x}, & (A4) \cr
a(x,2) & = a^{-\nu_x} \; b^{\mu_x}, & (A5)}
$$
where $\mu_x$ and $\nu_x$ are positive integers satisfying the following
recurrence relations
$$\eqalignno{
\mu_{x+1} &= 4\mu_x + \nu_x - \mu_{x-1}, & (A6) \cr
\nu_{x+1} &= 4\nu_x + \mu_x - \nu_{x-1}, & (A7)}
$$
with the initial conditions
$\mu_0 = \nu_0 = \nu_1 = 0$, and $\mu_1 = 1$. This can
easily be solved by defining the new functions
$$
m_x = \mu_x - \nu_x, \qquad n_x = \mu_x + \nu_x.
\eqno (A8)
$$
They satisfy uncoupled binary recurrence relations
$$
\eqalignno{
m_{x+1} & = 3m_x - m_{x-1}, & (A9) \cr
n_{x+1} & = 5n_x - n_{x-1}, & (A10) }
$$
with $m_0 = n_0 = 0$, $m_1 = n_1 = 1$. The solutions read
$$
\eqalignno{
m_x & = {1 \over \sqrt 5}\left[\left({3 + \sqrt 5 \over 2}\right)^x \,-\,
\left({3 - \sqrt 5 \over 2}\right)^x\right], & (A11) \cr
n_x & = {1 \over \sqrt {21}} \left[\left({5 + \sqrt {21} \over 2}\right)^x
\,-\, \left({5 - \sqrt {21} \over 2}\right)^x\right]. & (A12) }
$$
The first values of $m_x,\, n_x,\, \mu_x,\, \nu_x$ are given below in
Table II.

\bigskip
$$
\vbox{\offinterlineskip
\halign{&\vrule#&\strut\ #\ \cr
\multispan{21}\hfil\bf TABLE II\hfil\cr
\noalign{\medskip}
\noalign{\hrule}
height3pt&\omit&&\omit&&\omit&&\omit&&\omit&&\omit&&\omit&&\omit&&\omit&&\omit&\cr
&\hfil $x$\hfil&&\hfil 0\hfil&&\hfil 1\hfil&&\hfil 2
\hfil&&\hfil 3\hfil&&\hfil 4\hfil&&\hfil 5\hfil&&\hfil 6\hfil&&\hfil
7\hfil&&\hfil 8 \hfil&\cr
height3pt&\omit&&\omit&&\omit&&\omit&&\omit&&\omit&&\omit&&\omit&&\omit&&\omit&\cr
\noalign{\hrule}
height3pt&\omit&&\omit&&\omit&&\omit&&\omit&&\omit&&\omit&&\omit&&\omit&&\omit&\cr
&\hfil $m_x$\hfil&&\hfil 0\hfil&&\hfil 1\hfil&&\hfil 3
\hfil&&\hfil 8\hfil&&\hfil 21\hfil&&\hfil 55\hfil&&\hfil 144\hfil&&\hfil
377\hfil&&\hfil 987 \hfil&\cr
&\hfil $n_x$\hfil&&\hfil 0\hfil&&\hfil 1\hfil&&\hfil 5
\hfil&&\hfil 24\hfil&&\hfil 115\hfil&&\hfil 551\hfil&&\hfil $2\,640$\hfil&&
\hfil $12\,649$ \hfil&&\hfil $60\,605$ \hfil&\cr
&\hfil $\mu_x$\hfil&&\hfil 0\hfil&&\hfil 1\hfil&&\hfil 4
\hfil&&\hfil 16\hfil&&\hfil 68\hfil&&\hfil 303\hfil&&\hfil $1\,392$\hfil&&
\hfil $6\,513$\hfil&&\hfil $30\,796$ \hfil&\cr
&\hfil $\nu_x$\hfil&&\hfil 0\hfil&&\hfil 0\hfil&&\hfil 1
\hfil&&\hfil 8\hfil&&\hfil 47\hfil&&\hfil 248\hfil&&\hfil $1\,248$\hfil&&
\hfil $6\,136$\hfil&&\hfil $29\,809$ \hfil&\cr
height3pt&\omit&&\omit&&\omit&&\omit&&\omit&&\omit&&\omit&&\omit&&\omit&&\omit&\cr
\noalign{\hrule}
}}
$$

\bigskip
One may note that $m_x$ are precisely the even terms in the
standard Fibonacci sequence.

To determine the structure of $G$ in terms of the generators $a$ and
$b$, we note from ($A$3--5) that the only relations among them are
$$\eqalignno{
& a^{\mu_{L+1}} \; b^{-\nu_{L+1}} = I, &(A13) \cr
& a^{-\nu_{L+1}} \; b^{\mu_{L+1}} = I. &(A14)}
$$
According to the discussions in sections 3 and 4, the structure of $G$ is
given in terms of the elementary divisors $d_1,d_2$ of the matrix $\pmatrix{
\mu_{L+1} & -\nu_{L+1} \cr -\nu_{L+1} & \mu_{L+1}}$. We find that
$$
G_{L \times 2} = Z_{d_1} \times Z_{d_2}\,, \quad \hbox{with $d_1=(\mu^2_{L+1}-
\nu^2_{L+1})/d_2$ and $d_2={\rm gcd}(\mu_{L+1},\nu_{L+1})$}.
\eqno (A15)
$$
As a function of $L$, the number $d_2$ has an irregular behaviour, with very
large sudden jumps. The only result we could firmly establish follows from
the double inequality (4.12): $d_2 > 1$ if $L+1$ is divisible by 3, so in
that case, $G$ is not cyclic.

Rather surprisingly, the elementary divisors can equally be expressed in terms
of $m_{L+1}$ and $n_{L+1}$ by the neater formula
$$
d_1 = m_{L+1}n_{L+1}/d_2, \quad d_2={\rm gcd}(m_{L+1},n_{L+1}).
\eqno (A16)
$$
 From the relation ($A$8), Eqns ($A$15) and ($A$16) are clearly equivalent
when $m_x$ and $n_x$ are both odd, which is the case if $x$ is not divisible
by 3. That they are equivalent for $x$ a multiple of 3
lies in the delicate fact that the largest power of 2 that divides
$m_x$ is the same as that dividing $n_x$. To see this, we consider the
subsequences $M(r) = m_{3r}$ and $N(r) = n_{3r}$. We find from ($A$11--12)
that they satisfy the recursions
$$
\eqalignno{
M(r+1) & = 18\,M(r) - M(r-1), & (A17) \cr
N(r+1) & = 110\,N(r) - N(r-1), & (A18) }
$$
subjected to the initial conditions $M(0) = N(0) = 0$, $M(1) = 8$, and $N(1)
= 24$. Now the desired result that for each $r$, $M(r)$ and $N(r)$ have
the same 2--potency (i.e. the largest powers of 2 dividing them are the
same) follows from a more general result by the last author [22], which
gives a formula for the 2--potency for any `binary recursive sequence'
$\theta(r)$ defined by $\theta(0) = 0$, $\theta(1) =
\theta$, and $\theta(r+1) = 2R\,\theta(r) + S\,\theta(r-1)$, where $\theta$
is an arbitrary integer, and $R,S$ are odd integers.

\vskip 1truecm
\vfill \break
\noindent {\bf Appendix B}

\bigskip
In this appendix, we give the explicit formula giving the number $F(L)$ of
irreducible factors (over $Q$) of the polynomial $P_L(x)$ for arbitrary
$L$. In section 5, $P_L(x)$ was the characteristic equation of
the discrete Laplacian on a $L \times L$ lattice with open boundary
conditions and was defined by ($N=2(L+1)$)
$$
P_L (x) = \prod^L_{m,n=1} \left(x - 4
+ 2\cos {2\pi m \over N} + 2\cos {2\pi n \over N}\right).
\eqno (B1)
$$
The roots $\lambda_{m,n}=4-2\cos {2\pi m \over N}-2\cos {2\pi n \over N}$
of $P_L$ belong to the extension $Q(\cos{2\pi \over N})$ and
are permuted by its Galois group Gal$_L$, which acts on them by
(see section 5)
$$
\sigma_s(\lambda_{m,n}) \equiv \lambda_{sm,sn}, \quad
\hbox{for $s \in Z^*_N/\{\pm 1\}$}.
\eqno(B2)
$$

As discussed in section 5, finding the irreducible factors of $P_L$
amounts to split the set of roots $\{\lambda_{m,n}\}$ into orbits under
Gal$_L$, so that
$$
F(L) = |\{\lambda_{m,n} \;:\; 1 \leq m,n \leq L\}/{\rm Gal}_L|.
\eqno(B3)
$$
By a classical theorem (see for instance [23]), the number of orbits of
a set $X$ under the action of a group $G$ is equal to the average
number of fixed--points of $G$ in $X$, that is
$$
|X/G| = {1 \over |G|} \sum_{g \in G} |\{x \in X \;:\; gx=x \}|.
\eqno(B4)
$$
This allows to recast ($B$3) as
$$
F(L) = {2 \over \varphi(N)} \sum_{s \in Z^*_N/\{\pm 1\}} |\{1 \leq m,n \leq L
\;:\; \lambda_{sm,sn} = \lambda_{m,n} \}|.
\eqno(B5)
$$
 From the expression of $\lambda_{m,n}$, we have the obvious relations
$\lambda_{m,n} =
\lambda_{\pm m,\pm n} = \lambda_{\pm n,\pm m}$ where the signs are
uncorrelated. The only
subtle point of the analysis lies in the possible existence of other
relations. For example, if $L=4$ (i.e. $N=10$), $s$ takes the
two values 1 and 3, we have that $\sigma_3(\lambda_{1,4}) \equiv
\lambda_{3,2}$ is
actually equal to $\lambda_{1,4}=4$ even though $(3,2) \neq
(\pm 1,\pm 4)$ nor $(\pm 4,\pm 1) \bmod 10$. This situation is however not
generic, and can be proved to only happen for the pairs $(m,n)$ with
$m+n=L+1$ (related to the eigenvalue 4 of the Laplacian). Thus if we
leave them aside, we obtain the result that
$\lambda_{sm,sn}=\lambda_{m,n}$ if and
only if $(sm,sn) = (\pm m,\pm n)$ or $(\pm n,\pm m) \bmod N$. Each of the
$L$ eigenvalues $\lambda_{m,L+1-m}=4$ clearly forms an orbit on its
own under the Galois group and we obtain
$$\eqalign{
F(L) = L + {2 \over \varphi(N)} \sum_{s \in Z^*_N/\{\pm 1\}} |\{1 \leq
& m,n \leq L \;{\rm and} \; m+n \neq L+1 \;\;:\;\; \cr
& (sm,sn) = \pm (m,\pm n) \;{\rm or}\; \pm (n,\pm m) \bmod N \}|.}
\eqno(B6)
$$

The actual calculation of ($B$6) being straightforward but somewhat
lengthy, we only quote the final result.
Let $N = 2(L+1)$ have the prime factorization given by
$$
N = \prod_p p^{a_p}.
\eqno (B7)
$$
Then the number of irreducible polynomial factors of $P_L(x)$ equals
$$
\eqalign{
F(L) = & {N\over 2} + 6 + {1\over2} \prod_p {p^{a_p+1} + p^{a_p} - 2 \over
p - 1} + 2(2a_2 - 1) \prod_{p \not= 2} (2a_p + 1)  \cr
& + \prod_{p=1 \bmod 4} (2a_p +1) - (7a_2 + 5) \prod_{p \not= 2} (a_p + 1)
\cr
& + [\prod_{p \not= 2} (2a_p+1) - 1] \cdot \delta(a_2 = 1).}
\eqno (B8)
$$
It can be shown that the average growth of $F(L)$ is linear in $L$.
Some particular values are given in Table I, section 5.

\vskip 1truecm

\noindent {\bf Appendix C}

\bigskip
We list here the identity configurations $C_{\rm id}^{(L)}$, discussed in
sections 3 and 7, for $L = 4 \; {\rm and} \; 5$ and $L = 10 \; {\rm
and} \; 11$.

\medskip
$$
C_{\rm id}^{(4)}=\pmatrix{2&3&3&2 \cr 3&2&2&3 \cr 3&2&2&3 \cr 2&3&3&2}, \quad
C_{\rm id}^{(5)}=\pmatrix{2&3&2&3&2 \cr 3&2&1&2&3 \cr 2&1&0&1&2 \cr
3&2&1&2&3 \cr 2&3&2&3&2},
$$
$$
C_{\rm id}^{(10)}=\pmatrix{2&3&3&0&3&3&0&3&3&2 \cr 3&2&2&1&2&2&1&2&2&3 \cr
3&2&2&3&3&3&3&2&2&3 \cr 0&1&3&2&2&2&2&3&1&0 \cr 3&2&3&2&2&2&2&3&2&3 \cr
3&2&3&2&2&2&2&3&2&3 \cr 0&1&3&2&2&2&2&3&1&0 \cr
3&2&2&3&3&3&3&2&2&3 \cr 3&2&2&1&2&2&1&2&2&3 \cr 2&3&3&0&3&3&0&3&3&2}
$$
$$
C_{\rm id}^{(11)}=\pmatrix{2&3&3&0&3&2&3&0&3&3&2 \cr 3&2&2&1&2&1&2&1&2&2&3 \cr
3&2&2&3&3&2&3&3&2&2&3 \cr 0&1&3&2&2&1&2&2&3&1&0 \cr 3&2&3&2&2&1&2&2&3&2&3 \cr
2&1&2&1&1&0&1&1&2&1&2 \cr 3&2&3&2&2&1&2&2&3&2&3 \cr 0&1&3&2&2&1&2&2&3&1&0 \cr
3&2&2&3&3&2&3&3&2&2&3 \cr 3&2&2&1&2&1&2&1&2&2&3 \cr 2&3&3&0&3&2&3&0&3&3&2},
$$

\vskip 1truecm

\noindent {\bf Appendix D}

\bigskip
In [17], Lee and Tzeng have considered the following one--dimensional
deterministic sandpile model. The $L$ sites of a linear chain, labelled
0 to $L-1$, are assigned a height variable $z_i$ taking positive integer
values. A site $j$ becomes unstable if the difference $z_j - z_{j+1}$
exceeds some integer $N$, considered as a parameter. In this case, the
site $j$ looses $N$ grains of sand which fall on its $N$ right nearest
neighbours. They take open boundary conditions $z_i = 0$ for all $i \geq L$,
so that, under the toppling at site $j$, the sand is redistributed
according to $z_i \rightarrow z_i - \Delta_{ij}$, where the toppling matrix
is the lower triangular matrix
$$
\Delta _{ij} = \cases{N & if $i=j$, \cr
                      -1 & if $1 \leq i-j \leq N$. \cr}
\eqno(D1)
$$
The deterministic dynamics is defined by dropping, at each time step, one
grain of sand on the leftmost site $i=0$ and by letting the system
relax. The number of recurrent configurations is equal to $N^L$, also
equal to the determinant of $\Delta$. Note that this model is a
non--abelian sandpile model as the toppling condition is not local.

It was shown in [17] that for $N=2$ and 3, the recurrent configurations can be
labelled by the values of a toppling invariant, an integer function taken
modulo $N^L$, (see also [24]). They wrote down the invariant when $N=2$
for all values of $L$,
as well as when $N=3$ for some particular values of $L$. We show here that
our general method allows to prove that one invariant is sufficient to
label all recurrent configurations, for all $N$, and we will compute this
invariant for $N=2,3$, all values of $L$.

That only one invariant gives a complete labelling of the recurrent
configurations amounts to prove that the elementary divisors of $\Delta$
are $d_1=N^L$ and $d_i=1$ for $i \geq 2$. Equivalently, it amounts to
show that among all the invariants (3.3) constructed from $\Delta ^{-1}$,
only one is independent. The matrix $\Delta^{-1}$ is easily obtained
and read
$$
\Delta^{-1}_{ij} = \sum^{L-1}_{k=0} \, c_k \, \delta_{i-j,k},
\eqno(D2)
$$
where the entries $c_k$ satisfy the recurrence relations
$$
\eqalign{
& Nc_k = c_{k-1} + c_{k-2} + \ldots + c_{k-N}, \cr
& c_0 = {1 \over N}, \quad c_{-k}=0 \;\hbox{ for all $k \geq 1$}. \cr}
\eqno(D3)
$$
We thus obtain the $L$ invariants
$$
Q_i(C) = \sum_{k=0}^i \, c_{i-k}z_k \; \bmod 1, \quad i=0,1,\ldots L-1.
\eqno(D4)
$$
The first invariants read
$Q_0 = {1 \over N}z_0 \bmod 1$, $Q_1 = {1 \over N^2}z_0 + {1 \over N}z_1
\bmod 1$, $Q_2 = {N+1 \over N^3}z_0 + {1 \over N^2}z_1 + {1 \over N}z_2
\bmod 1$, so that we have
$$\eqalignno{
& NQ_1 = Q_0 \bmod 1, &(D6a) \cr
& NQ_1 = (N+1)Q_0 \bmod 1, \qquad NQ_2 = (N+1)Q_1 \bmod 1, &(D6b) \cr}
$$
showing that $Q_0$ can be expressed in terms of $Q_1$, itself expressable
in terms of $Q_2$. It is not difficult to show recursively that this pattern
continues to hold, namely there exists, for all $i$ between 1 and $L-1$, an
integer $X_i$ coprime with $N$ such that
$$
NQ_j = X_i Q_{j-1} \bmod 1, \quad \hbox{for all $1 \leq j \leq i$}.
\eqno(D7)
$$
(From (D6) we have $X_1=1$ and $X_2=N+1$.) Equation (D7) implies that
all invariants $Q_0,Q_1,\ldots,Q_{L-2}$ can be expressed in terms of
$Q_{L-1}$ which is thus the only independent one and which provides a
complete labelling of the set of recurrent configurations.

To compute the invariant $Q_{L-1}$, we note that the coefficients $c_k$
satisfy in fact a recurrence relation of order $N-1$:
$$
Nc_k + (N-1)c_{k-1} + \ldots + c_{k+1-N} = Nc_{N-1} + (N-1)c_{N-2} +
\ldots + c_0 = 1.
\eqno(D8)
$$
For $N=2$ and 3, the solutions to (D8) read ($k \geq 0$)
$$\eqalignno{
N=2 \;\;&:\;\; c_k = {1 \over 3}[1-({-1 \over 2})^{k+1}], &(D9a) \cr
N=3 \;\;&:\;\; c_k = {1 \over 6} + {1 \over 12} \cdot \left[
\left({-1-\sqrt{-2}\over 3}\right)^k + \left({-1+\sqrt{-2}\over 3}
\right)^k \right].
&(D9b) } $$
Using these values in $Q_{L-1}$ of (D4) yields the desired result. For
higher values of $N$, the coefficients $c_k$ are given by the
generating function
$$f_N(t) = \sum^\infty_{k=0} \, c_k^{(N)}t^k = (1-t)\,.\, \left[t^{N+1}
- (N+1)t + N \right]^{-1}.
\eqno(D10)
$$

\vfill
\eject
\noindent{\bf References}

\bigskip
\item{[1]}
P. Bak, C. Tang and K. Wiesenfeld, {\it Self-organised criticality :
an explanation of 1/f noise}, Phys. Rev. Lett. 59 (1987)
381-384.

\item{[2]}
P. Bak, C. Tang and K. Wiesenfeld, {\it Self-organised criticality}
Phys. Rev. A38 (1988) 364-374.

\item{[3]}
P. Bak and K. Chen, {\it The physics of fractals}, Physica D38 (1989)
5-12.

\item{[4]}
K. Chen, P. Bak and S.P. Obukhov, {\it Self-organised criticality in a
crack propogation model of earthquakes}, Phys. Rev. A43 (1991)
625-630.

\item{[5]}
A. Sornette and D. Sornette, {\it Self-organised criticality of
earthquakes} Europhys. Lett. 9 (1989) 197-202.

\item{[6]}
D. Dhar, {\it Self-organised critical state of sandpile automaton
models}, Phys. Rev. Lett. 64 (1990) 1613-1616.

\item{[7]}
D. Dhar and R. Ramaswamy, {\it An exactly solved model of
self-organised critical phenomena}, Phys. Rev. Lett. 63 (1989) 1659-1662.

\item{[8]}
S.N. Majumdar and D. Dhar, {\it Equivalence between the abelian
sandpile model and the q $\rightarrow$ 0 limit of the Potts model},
Physica A185 (1992) 129-145.

\item{[9]}
D. Dhar and S.N. Majumdar, {\it Abelian sandpile model on Bethe
lattice}, J. Phys. A23 (1990) 4333-4350.

\item{[10]}
B. Gaveau and L.S. Schulman, {\it Mean-field self-organised
criticality}, J. Phys. A: Math. Gen. 24 (1991) L475-480.

\item{[11]}
S.A. Janowski and C.A. Laberge, {\it Exact solution for a mean-field
abelian sandpile}, J. Phys. A: Math. Gen. 26 (1993) L973-980.

\item{[12]}
D. Dhar and S.S. Manna, {\it Inverse avalanches in the abelian
sandpile models}, Phys. Rev. E49 (1994) 2684-2687.

\item{[13]}
M. Creutz, {\it Abelian sandpiles}, Computers in Physics 5 (1991) 198-203.

\item{[14]}
S.H. Liu, T. Kaplan and L.J. Gray, {\it Geometry and dynamics of
deterministic sandpiles}, Phys. Rev. A42 (1990) 3207-3212.

\item{[15]}
K. Wiesenfeld, J. Theiler and B. McNamara, {\it Self-organised
criticality in a deterministic automaton}, Phys. Rev. Lett. 65 (1990) 949-952.

\item{[16]}
M. Markosova and P. Markos, {\it Analytical calculation of the
attractor period of deterministic sandpile}, Phys. Rev. A46 (1992)
3531-3537.

\item{[17]}
S.C. Lee and W.J. Tzeng, {\it Hidden conservation law for sandpile
models} Phys. Rev. A45 (1992) 1253-1254.

\item{[18]}
N. Jacobson, {\it Basic Algebra} (W.H. Freeman, San Fransisco, 1974).

\item{[19]}
H. Cohen, {\it A Course in Computational Algebraic Number Theory}
(Springer, Berlin, Heidelberg, 1993).

\item{[20]}
Ian Stewart, {\it Galois Theory} (Chapman and Hall, London, New York 1973).

\item{[21]}
P. Bak and M. Creutz, 1993, BNL preprint 48309.

\item{[22]}
D.N. Verma, unpublished.

\item{[23]}
M. Armstrong, {\it Groups and Symmetry} (Springer, Berlin, Heidelberg,
New York, 1988).

\item{[24]}
E.R. Speer, {\it Asymmetric abelian sandpile models}, J. Stat Phys. 71
(1993) 61-74.

\end

\bye